# Atomic-scale control of magnetic anisotropy via novel spin-orbit coupling effect in La$_{2/3}$Sr$_{1/3}$MnO$_3$/SrIrO$_3$ superlattices

Di Yi[‡][*][1], Jian Liu[‡][*][2,3,4], Shang-Lin Hsu[1,5], Lipeng Zhang[7], Yongseong Choi[6], Jong-Woo Kim[6], Zuhuang Chen[1], James D. Clarkson[1], Claudy R. Serrao[1], , Elke Arenholz[8], Philip J. Ryan[6], Haixuan Xu[7], Robert J. Birgeneau[1,3,4] and Ramamoorthy Ramesh[1,3,4]

1. *Department of Materials Science and Engineering, University of California, Berkeley, California 94720, USA*
2. *Department of Physics and Astronomy, University of Tennessee, Knoxville, Tennessee 37996, USA*
3. *Department of Physics, University of California, Berkeley, California 94720, USA*
4. *Materials Science Division, Lawrence Berkeley National Laboratory, Berkeley, California 94720, USA*
5. *National Center for Electron Microscopy, Lawrence Berkeley National Laboratory, Berkeley, California 94720, USA*
6. *Advanced Photon Source, Argonne National Laboratory, Argonne, Illinois 60439, USA*
7. *Department of Materials Science and Engineering, University of Tennessee, Knoxville, Tennessee 37996, USA*
8. *Advanced Light Source, Lawrence Berkeley National Laboratory, Berkeley, California 94720, USA*

* Email: yid@berkeley.edu, jianliu@utk.edu

[‡] These authors contributed equally to this work

# Abstract

Magnetic anisotropy (MA) is one of the most important material properties for modern spintronic devices. Conventional manipulation of the intrinsic MA, i.e. magnetocrystalline anisotropy (MCA), typically depends upon crystal symmetry. Extrinsic control over the MA is usually achieved by introducing shape anisotropy or exchange bias from another magnetically ordered material. Here we demonstrate a pathway to manipulate MA of *3d* transition metal oxides (TMOs) by digitally inserting non-magnetic *5d* TMOs with pronounced spin-orbit coupling (SOC). High quality superlattices comprised of ferromagnetic $La_{2/3}Sr_{1/3}MnO_3$ (LSMO) and paramagnetic $SrIrO_3$ (SIO) are synthesized with the precise control of thickness at atomic scale. Magnetic easy axis reorientation is observed by controlling the dimensionality of SIO, mediated through the emergence of a novel spin-orbit state within the nominally paramagnetic SIO.

Magnetic anisotropy (MA) is one of the fundamental properties of magnetic materials. The widespread scientific interest in MA originates from its decisive role in determining a rich spectrum of physical responses, such as the Kondo effect (1), the magneto-caloric effect (2), magnetic skyrmions (3), etc. From a technological viewpoint, it is an important and promising approach to control MA by external stimuli, such as electric field (4). In general, there are two approaches to design MA of a ferromagnet. In the first approach one manipulates the intrinsic magnetocrystalline anisotropy (MCA), deriving from the local crystal symmetry and spin-orbit coupling (SOC) of the magnetic ion (5-7). Alternatively, one can tune the MA though extrinsic contributions to the anisotropy such as shape (8) or exchange coupling to a strong antiferromagnet (9).

One focus of magnetism research is $3d$ transition metal oxides (TMOs), a class of materials that exhibit various functionalities including ferromagnetism due to the strong electron-electron correlation. However SOC is usually weak or negligible in $3d$ TMOs. On the other hand, the pronounced SOC of heavy elements has drawn attention in recent years due to the emergence of new topological states of matter (10-12) and spintronics (13, 14). In contrast to $3d$ TMOs, the correlation strength is often too small in $5d$ TMOs to host magnetism. Therefore it is an interesting approach to design systems that combine the merits of these two fundamental interactions. Similar ideal has been studied in metal multilayers (15, 16). However it still remains an important challenge to explore the ideal in complex oxides, where a variety of emergent phenomena have been discovered due to the power of atomic-scale confinement and interfacial coupling (17-21).

Here we present an approach towards accomplishing this goal by atomic-scale synthesis. By fabricating high quality superlattices comprised of $3d$ and 5d TMOs, we address two open questions: the effect of SOC on the functionality of $3d$ TMOs and the possible emergent magnetic state of $5d$ TMOs. So far this approach has been limited and overlooked. To the best of our knowledge, SrTiO$_3$/SrIrO$_3$ is the only $3d$/$5d$ superlattice that has been experimentally studied (22), which reveals the effect of dimensional confinement. However the 3$d$ state is rather inactive in that system. Here we study a model system comprised of ferromagnetic La$_{2/3}$Sr$_{1/3}$MnO$_3$ (LSMO) and paramagnetic SrIrO$_3$ (SIO). We have discovered that the magnetic easy axis of LSMO rotates between two crystallographic directions, i.e. ⟨100⟩ and ⟨110⟩ (pseudo-cubic) by digitally reducing the SIO thickness down to one monolayer. Remarkably, the reorientation of MA is accompanied by the emergence of a large spontaneous, orbital-dominated magnetic moment of the 5$d$ electrons, revealing a heretofore-unreported spin-orbit coupled state.

## Results and Discussion

The colossal magnetoresistive system LSMO is a 3$d$ ferromagnet with a high Curie temperature (23). Due to the potential for applications in all-oxide spintronics, the MA of LSMO thin films has been investigated extensively. Previous studies have established the magnetic easy axis of an epitaxial LSMO thin film on (001) oriented SrTiO3 (STO) substrate to be in-plane along the crystallographic ⟨110⟩ (7, 24) as a consequence of the strain superimposed on the intrinsic rhombohedral symmetry. SIO is an end member of the Ruddlesden-Popper (RP) series Sr$_{n+1}$Ir$_n$O$_{3n+1}$(25). It has been identified to be a spin-orbit-coupled paramagnet without any signature of long-range magnetic ordering (26, 27), likely due to the topological nature of its metallicity (28). Owing to the structural compatibility, it has been theoretically proposed as a key building block for stabilizing topological phases at interfaces and in superlattices (12, 29). To investigate the impact of artificial confinement and interfacial coupling, we intentionally insert $m$ unit cells (uc) of SIO every 3uc of LSMO with $m$ being varied from 10 to 1 in order to scale down the SIO layer from 4nm to 0.4nm (labeled as SL3m). All the superlattices are deposited on SrTiO3 (STO) (001) substrates. The precise control of thickness is achieved by monitoring the intensity oscillations of the reflection high-energy electron diffraction (RHEED) pattern during the growth (SI Appendix, Fig. S1), revealing the layer-by-layer growth mode of both LSMO and SIO. This growth mode is critical for synthesizing high quality superstructures. Details of the synthesis protocols used in our study are provided in the Methods section.

The high quality of the superlattices characterized by several techniques demonstrates the precise control of thickness at the atomic-scale. Fig. 1A shows a scanning transmission electron microscopy (STEM) image of a superlattice with differing periodicities in repeated patterns. The sharp Z-contrast of B-site species across the interface, supplemented by the line profile of the electron energy loss spectroscopy (EELS) of A-site La atoms (SI Appendix, Fig. S2), indicates minimal inter-diffusion at the interface. Fig. 1B shows the high-resolution θ-2θ x-ray diffraction of the SL31 and SL35. The satellite peaks corresponding to the superlattice structure and the finite size oscillations arising from the thickness are pronounced, suggesting the high degree of interface abruptness and agreement with the intended periodicity. Fig. 1C shows the reciprocal spacing mapping (RSM) of sample SL35, revealing that the superlattice is coherently strained by the STO substrate. Further structural characterization data are shown in SI Appendix, Fig. S1.

The temperature-dependence of the magnetization of the superlattices (SI Appendix, Fig. S4) is similar to that of pure LSMO thin film (albeit with a decrease of $T_c$ as $m$ increases), indicating that the overall magnetization is dominated by the ferromagnetic LSMO component. In order to study the MA, magnetization loops are measured along different crystallographic axes of SL3$m$ (Materials and Methods section). First, the magnetic easy axis is revealed to be in the film plane by comparing the in-plane and out-of-plane magnetic loops (SI Appendix, Fig. S4), consistent with our expectations (due to the strain effect and shape anisotropy). Additionally, magnetization loops along symmetry-equivalent in-plane directions, e.g. [100] and [010] (Fig. 2B), demonstrate that the MA is biaxial (four-fold rotational symmetry with $\pi/2$ periodicity), which is indicative of MCA and thus rules out the influence of shape anisotropy (8).

The impact of SIO is demonstrated by the systematic influence of the SIO layer thickness on the MA (Fig. 2A), which is represented by the normalized difference between the remnant magnetization along ⟨100⟩ and ⟨110⟩ directions. The positive sign corresponds to the ⟨100⟩ easy axis while the negative sign indicates a $\pi/4$ shift to ⟨110⟩. As can be seen, the superlattices with long periodicity ($m>5$, i.e. 2nm) exhibit ⟨110⟩ easy axis, identical to that of pure LSMO thin film (purple dot, Fig. 2A). Intriguingly, as $m$ reduces ($m<5$), a reorientation of the easy axis to ⟨100⟩ is observed. The magnitude of the normalized difference systematically increases as $m$ decreases, revealing a tunability of ~40% (theoretical limit ~58% of the biaxial MA, magnetic moments aligning along one direction have a $\pi/4$ projection on the other direction). We also carried out anisotropic magnetoresistance (AMR) measurements to validate the observed MA. The longitudinal resistance is measured along ⟨100⟩ direction and the magnetic field is rotated in-plane with respect to the current direction (Materials and Methods section). Fig. 2C shows the polar plots of AMR of SL33 and SL310. The four-fold rotation of AMR reflects the same symmetry of the MCA. The $\pi/4$ phase shift of AMR between SL33 and SL310 is coincident with the MA evolution shown in Fig. 2A. Further analysis of the AMR of SL3$m$ is shown in SI Appendix Fig. S5 and confirms the change of MA as the SIO thickness reduces. The temperature dependence of the novel MA (⟨100⟩ easy axis) is acquired by measuring magnetic hysteresis loops in different orientations at multiple temperatures. Fig. 3A shows that the MA persists to the Curie temperature (~270K) of the superlattice.

A close examination of the results discussed above reveals the unique nature of the MA tailoring. First, as pointed out, the MA with a $\pi/4$ phase shift of easy axis is not due to the shape anisotropy (8). Since the LSMO dominates the aggregate magnetization, one must consider the potential contribution from LSMO crystal symmetry change. Previous studies have revealed a possible mechanism that could lead to the reorientation of in-plane easy axis of LSMO thin films. It has been demonstrated that a moderate biaxial compressive strain on LSMO could lead to the orthorhombic structure and ⟨100⟩ easy axis due to the asymmetry of octahedral rotation patterns (5). RSM measurements (represented by SL35 in Fig. 1C) reveal that our superlattices are coherently constrained by the substrate, confirming that LSMO is under biaxial tensile strain. Another possible contribution is the interfacial octahedral coupling (30), considering the difference between LSMO and SIO (23, 26). In this scenario, one expects the rotational pattern of the LSMO to be unaltered by the thinner SIO (30), therefore the short period superlattice (SL31) would have a reduced tendency than the long period superlattice (SL310) to show the reorientation of magnetic easy axis compared to pure LSMO film. This is however opposite to the observed thickness evolution in Fig. 2A. In fact, x-ray diffraction measurements of several half ordering reflections of SL31 rule out the alteration of octahedral rotation pattern as the origin (SI Appendix section b and Fig. S3). The results thus imply a distinct role of the strong SOC in SIO to engineer the MA.

To gain more insight in the spin-orbit interaction, we investigated the valence and magnetic state of both LSMO and SIO by carrying out element selective X-ray absorption spectroscopy (XAS) and X-ray magnetic circular dichroism (XMCD) measurements (31, 32). Fig. 3C and D show the XAS spectra of Mn (red curve) and Ir (blue curve) of the SL31 taken at the resonant $L_{2,3}$ edges. As a comparison, XAS spectra of reference samples of LSMO (purple curve, Fig. 3C) and SIO (purple curve, Fig. 3D) thin films were taken simultaneously. The absence of peak position shift and the identical multiplet features suggest a minimal effect of charge transfer between Mn and Ir cations. Figure 3E shows the Mn and Ir XMCD spectra. The large dichroism at Mn edge is expected for the highly spin-polarized ferromagnetic LSMO and consistent with magnetometry. However the presence of a sizable XMCD at the Ir edge reveals the onset of a net magnetization, unexpected for the paramagnetic SIO (26). To validate this observation, the XMCD spectra were taken by multiple measurements with alternated x-ray helicity and magnetic field (Materials and Methods section). The opposite sign of dichroism of the two cations indicates that the Mn and Ir net moments are antiparallel to each other. This nontrivial coupling is further demonstrated by the coincident reversal of LSMO magnetization and Ir-XMCD (Fig. 3B). Furthermore, the temperature dependence of Ir edge XMCD (Fig. 3A) reveals a relatively high onset temperature (near room temperature), which is closely related to the Curie temperature of LSMO. The combination of these results suggests the emergence of magnetic ordering in the nominally paramagnetic SIO in the ultrathin limit.

To understand the origins of the Ir moments, sum-rules analysis of XMCD spectra in Fig. 3E was applied to differentiate the spin component from the orbital counterpart (33), which yields an unexpected result. A relatively large orbital moment $\boldsymbol{m_l = (0.036 \pm 0.003)\ \mu_b/Ir}$ is obtained for SL31 compared with the effective spin component $\boldsymbol{m_{se} = (0.002 \pm 0.003)\ \mu_b/Ir}$ (SI Appendix section e and Fig. S6). Such a large ratio of $\boldsymbol{m_l/m_{se}}$ is to date unreported even in the 5d TMOs. As a comparison, sum rules analysis was also applied to the Mn L edge, which yields a $\boldsymbol{m_l/m_{se}}$ ratio less than 0.01 and is consistent with the dominant role of the spin moment for 3d TMOs (SI Appendix section e and Fig. S7).

In order to further understand the magnetic behavior of SIO within the confines of the superstructure environment, we also performed first-principles density functional calculations with generalized gradient approximation (GGA) + Hubbard U + spin-orbit coupling (SOC) (SI Appendix section f). We compared the energies of configurations where the Mn moments align in the ⟨100⟩ and ⟨110⟩ directions in SL31. Since correlated oxides are notoriously challenging for GGA, we explored a variety of U parameter combinations. While the magnitude of the energy difference depends on the details of the parameters, the ⟨100⟩ direction is energetically more favorable than the ⟨110⟩ direction in SL31, consistent with the experiments. Moreover, the monolayer of SIO in SL31 develops a canted in-plane antiferromagnetic ordering (weak ferromagnetism), which is similar to the magnetic ordering of $Sr_2IrO_4$ (34). However, while the moments of $Sr_2IrO_4$ are known to align along ⟨110⟩ direction (35, 36), the moments of SIO in SL31 prefer ⟨100⟩, highlighting a key distinction in terms of MA (Fig. S8). In summary, XMCD and first-principles calculations both reveal the emergence of the weak ferromagnetism in the ultrathin SIO, which shows an orbital-dominated moment and a different magnetic anisotropy compared to the Ruddlesden-Popper series iridates (discussed in SI Appendix section g).

This distinctive character of the Ir moment in the superlattices presents an unconventional spin-orbit state in the iridate family. Due to the strong spin-orbit coupling, the low-spin $d^5$ configuration of $Ir^{4+}$ valence state within the octahedral crystal field fills the *d*-shell up to half of a spin-orbit coupled doublet ($J_{eff}=1/2$ state), which has been theoretically established (37) and experimentally observed (34) for several iridates, for example, $Sr_2IrO_4$. This $J_{eff}=1/2$ state, regarded as the main driver of Ir related physics, is characterized by the distinct orbital character and orbital mixing of $t_{2g}$ bands, leading to the ratio of $\boldsymbol{m_l/m_{se}}$ ~0.5 (37). Experimentally it is characterized by the absence of the $L_2$ edge magnetic dichroism, due to dipole selection rules (34, 38). Our XMCD results of SL31 unambiguously reveal the breakdown of the $J_{eff}=1/2$ picture in the superlattices by considering how the sign and amplitude of $L_2$ and $L_3$ edge XMCD signatures are equivalent and comparable respectively, thereby yielding a large $\boldsymbol{m_l/m_{se}}$ ratio as discussed above. In order to enhance the orbital component relative to the spin component, the new spin-orbit state is likely to be formed by mixing the $J_{eff}=1/2$ state with the $J_{eff}=3/2$ state, where the two components are antiparallel (discussed in SI Appendix section e). Moreover this spin-orbit state was not reported before in the STO/SIO example (22) that is dominated by dimensional confinement, which also clearly implies the decisive role of interfacial coupling, beyond the dimensionality effects.

The emergent weak ferromagnetism in SIO exhibits a close correlation to the control of MA. As the thickness *m* reduces, the stability of ⟨100⟩ easy axis (Fig. 2A), the emergent weak ferromagnetism (Fig. 3E) and the $m_l/m_{se}$ ratio (SI Appendix section e, Fig S6) become more significant. In addition, the emergent weak ferromagnetism shows a similar temperature dependence to that of the MA with <100> easy axis (Fig. 3A). Thus the results suggest a crucial contribution to the overall MA from the interfacial magnetic coupling between the Mn spin moments and the emergent Ir orbital moments. The effective in-plane biaxial anisotropic energy is commonly defined as: $E = K_{eff}M^4\sin^2 2\theta$ (39), where M is the in-plane magnetization and $\theta$ is the angle to $\langle 100 \rangle$ (Fig. 4A). Taking into account the superlattice geometry, the effective anisotropy $K_{eff}$ is determined by the competition between MCA of LSMO ($K_c$) and SIO-induced anisotropy ($K_{in}$) (Fig. 4D)). The sign of $K_c$ remains negative due to the absence of structure change discussed before, which favors the $\langle 110 \rangle$ easy axis (Fig. 4A). The sign of $K_{in}$ is positive which favors the $\langle 100 \rangle$ easy axis (Fig. 4B). Therefore in the short-period superlattices where the emergent weak ferromagnetism is more significant, $K_{in}$ overcomes $K_c$ and becomes dominant in $K_{eff}$ (Fig. 4C). As $m$ increases digitally, $K_{eff}$ evolves from positive ($K_{in}$ dominated) to negative ($K_c$ dominated), manifesting itself as a systematic evolution of MA (Fig. 4E). The sign of $K_{in}$ reflects the magnetic anisotropy of the emergent Ir moment that the Mn couples to. As discussed above, in contrast to the $J_{eff}=1/2$ RP phases $Sr_2IrO_4$ (34) and $Sr_3Ir_2O_7$ (40), the new spin-orbit state in the superlattices features as a mixture of $J_{eff}=1/2$ and $J_{eff}=3/2$. Unlike $J_{eff}=1/2$ which is actually an atomic $J=5/2$ spin-orbit state, $J_{eff}=3/2$ is not an eigenstate of the SOC operator albeit being an eigenstate of the octahedral crystal field operator. As a result, the $J_{eff}=3/2$ wavefunctions are further hybridized with $e_g$ orbitals and their energies acquire corrections proportional to the ratio of SOC and crystal field. Therefore the crystal field tends to lock the total angular moment along its principle axis, e.g. $\langle 100 \rangle$, leading to a large single-ion anisotropy which is absent in $J_{eff}=1/2$ (discussed in SI Appendix section g). Thus the mixture of $J_{eff}=1/2$ and $J_{eff}=3/2$, which can be engineered in the superlattices, controls the magnetic anisotropy of the emergent Ir moment. This result proffers a new control paradigm in correlated electron behavior.

In conclusion, we present the ability to engineer the MA of ferromagnetic LSMO by inserting the strong SOC paramagnet SIO with atomically controlled thickness. The origin is attributed to a novel spin-orbit coupled state with a relatively large orbital-dominated moment that develops in the typically paramagnetic SIO. Our results demonstrate the potential of combined artificial confinement and interfacial coupling to discover new phases as well as to control the functionalities. This study particularly expands the current research interest of the atomic-scale engineering towards the strong SOC *5d* TMOs, which also paves the way towards all-oxide spintronics.

## Methods

**Synthesis** $(LSMO)_3(SIO)_m$ superlattices with different *m* were grown by RHEED-assisted Pulsed Laser Disposition on low-miscut STO substrates. Before the growth, the substrates were wet-etched by buffered HF acid, followed by a thermal annealing process at 1000 °C for 3 hours in oxygen atmosphere. Both LSMO and SIO sublayers were deposited at 700 °C and 150 mTorr oxygen partial pressure from the chemical stoichiometric ceramic target by using the KrF excimer laser (248nm) at the energy density of 1.5J/cm$^2$. The repetition rate was 1Hz and 10Hz for each sublayers. During the growth, in-situ RHEED intensity oscillations were monitored to control the growth at the atomic scale. After growth the samples cooled down at the rate of 5 °C/min in pure oxygen atmosphere.

**Magnetic and transport measurement** Magnetic measurements were performed on the Quantum Design SQUID magnetometry with an RSO option, which provides a sensitivity of $10^{-7}$ emu. In order to study the MA, magnetization loops were measured along different crystallographic directions of SL3*m*: in-plane [100], in-plane [110] and out-of-plane [001]. Also magnetization loops were measured along symmetry-equivalent in-plane directions, e.g. [100] and [010], to check the angle dependence of MA. Transport measurements were performed using the Quantum Design Physical Property Measurement System (PPMS, 14T). The longitudinal resistance is measured by the four-point probes method with the excitation current (10 µA) flowing in the film plane (crystallographic [100], shown in Fig. 2 (c)). The relative angle between the magnetic field and current was controlled by rotating the sample holder. The magnetic hysteresis loops and AMR curves reported here have been reproduced on multiple samples.

**XAS, XMCD and XRD measurement** The XAS and XMCD characterization at the Mn edge were carried out at beamline 4.0.2 at the Advanced Light Source, Lawrence Berkeley National Lab. The measurements were performed using the total-electron-yield (TEY) mode and the angle of incident beam is $30°$ to the sample surface. The XAS and XMCD characterizations at Ir edge were carried at beamline 4-ID-D at APS in Argonne National Lab. The results were taken by collecting the fluorescence yield signal and the incident beam is $3°$ to the sample surface. All of the XMCD spectra were measured both in remanence and in saturation field. Experimental artifacts were ruled out by changing both the photon helicity and the magnetic field direction. Since the XMCD spectra of the Ir-edge is relatively weak, multiple measurements were repeated to increase the signal-to-noise ratio of the spectra (5 times for each spectrum at each field). Also we measured the spectra at different times and on different samples. The hysteresis loop of the Ir-XMCD was measured with energy fixed at 12.828 keV (maximum of $L_2$ XMCD) by altering the photon helicity at each magnetic field. Synchrotron XRD measurements were carried out at sector 33BM and 6-ID-B at APS in Argonne National Lab.

## Acknowledgement

We acknowledge Dr. Daniel Haskel, E. Karapetrova, S. Cheema, and Dr. J.H. Chu and for experiment assistance and Professor L.W. Martin, Professor M. van Veenendaal and Professor E. Dagotto for critical discussions. This work was funded by the Director, Office of Science, Office of Basic Energy Sciences, Materials Science and Engineering Department of the U.S. Department of Energy (DOE) in the Quantum Materials Program (KC2202) under Contract No. DE-AC02-05CH11231. D.Y. is supported by the National Science Foundation (NSF) through Materials Research Science and Engineering Centers (MRSEC) Grant No. DMR 1420620. We acknowledge additional support for the research at UC Berkeley through the U.S. Department of Defense (DOD) Army Research Office (ARO) Multidisciplinary University Research Initiatives (MURI) Program, Defense Advanced Research Projects Agency (DARPA) and Center for Energy Efficient Electronics Science (E3S). Research at the University of Tennessee (J.L., L. Z. and H.X) is sponsored by the Science Alliance Joint Directed Research and Development Program. This research used resource of The National Institute for Computational Sciences (NICS) at the University of Tennessee under contract UT-TENN0112. This research used resources of the Advanced Photon Source, a U.S. Department of Energy (DOE) Office of Science User Facility operated for the DOE Office of Science by Argonne National Laboratory under Contract No. DE-AC02-06CH11357. Use of the Advanced Light Source is supported by the Director, Office of Science, Office of Basic Energy Sciences, of the U.S. Department of Energy under Contract No. DE-AC02-05CH11231. Work at the National Center for Electron Microscopy, Molecular Foundry is supported by the Office of Science, Office of Basic Energy Sciences, of the U.S. Department of Energy under Contract No. DE-AC02-05CH11231.

## Author Contributions

D.Y and J.L. designed, with R.B and R.R., directed the study, analyzed results, and wrote the manuscript. S.L.H. performed the STEM measurement and provided analysis. Z.C., J.C., P.R. and J.K aided in the structure characterizations. The XAS and XMCD measurements were aided by Y.C. and E.A. L.Z. and H.X. provided the theoretical support. All authors made contributions to write the manuscript.

**Figure Legends:**

**Fig.1 Structural characterization of the LSMO/SIO superlattice.** (A) High-angle annular dark-field (HAADF) STEM images of LSMO/SIO superlattice with designed periodicities in one sample. The four high magnification images correspond to the regions of $(LSMO)_1/(SIO)_1$, $(LSMO)_2/(SIO)_2$, $(LSMO)_3/(SIO)_3$ and $(LSMO)_5/(SIO)_5$ from top to bottom (the number refers to the thickness in unit cell). The atoms are marked by different colors: Ir (brightest contrast) in orange, Mn (darkest contrast) in green, and A-site atoms in blue. (B) θ-2θ X-ray diffractograms of a SL31 (top) and a SL35 (bottom) superlattice. Both the superlattice peaks and the thickness fringes reveal the high degree of interface abruptness. (C) X-ray reciprocal spacing mapping of a SL35 superlattice around (103) peak, confirming coherent growth of the superlattice.

**Fig.2 Magnetic and transport characterization of the LSMO/SIO superlattice.** (A) Dependence of magnetic anisotropy (MA) on SIO thickness ($m$) in the superlattice series SL3m. MA is defined as the difference of remnant moment MR between two crystallographic directions normalized by the saturation moment MS (($M_R[100] - M_R[110])/M_S$). The positive sign corresponds to the ⟨100⟩ easy axis while the negative sign indicates a $\pi/4$ shift to ⟨110⟩. The purple dot shows the anisotropy of LSMO thin film. Magnetic easy axis is shown by the arrow in the oxygen octahedral for the series of superlattices. Error bars are derived from measurements on multiple samples. (B) Magnetic hysteresis loops of a SL31 superlattice with magnetic field H in [100] (blue), [110] (red) and [010] (black) crystallographic orientation. The inset is the plot of magnetic hysteresis loops of LSMO (20nm) thin film on STO as a comparison. Magnetization is averaged by LSMO thickness in this study. (C) Schematic and polar plots of in-plane anisotropic magnetoresistance (AMR). The current is along [100] direction and the magnetic

field (1T) is rotated within the film plane. The polar plots show a phase shift of π/4 between SL33 and SL310, consistent with the thickness evolution of MA in (A).

**Fig.3 XAS and XMCD spectra of the LSMO/SIO superlattice.** (A) Temperature dependence of the magnetization, MA and the XMCD (Ir edge) of SL31. (B) Field dependence of magnetization measured by magnetometer and sign of Ir-XMCD. The opposite sign corresponds to the antiparallel configuration of Mn and Ir moments in (E). (C), (D) Core-level XAS spectra of Mn and Ir of the superlattice SL31 along with the spectra of SIO and LSMO thin film (purple curve). Peak positions of the XAS spectra of the superlattice are same as the pure thin films and the multiplet features are identical in both Ir and Mn edge within the experimental limit, suggesting the minimal effect of charge transfer in the superlattice. (E) XMCD spectra of the multiple superlattices measured at 10K with 1T applied along [100] direction with the same photon helicity and field direction. The magnitude of XMCD is normalized by the magnitude of $L_3$ XAS for Mn and Ir respectively. For comparison, the magnitude of XMCD of Ir is multiplied by a factor of 25.

**Fig.4 Schematic diagram of the origin of the MA engineering.** (A) Magnetic anisotropy energy of LSMO. The anisotropy energy is defined by the formula: $E = K \cdot M^4 \sin^2 2\theta$, whereas M is the in-plane magnetization and $\theta$ is the angle to [100] (red arrows). The MCA of LSMO ($K_c$ <0) favors ⟨110⟩ easy axis as shown by the black solid lines (energy minimum). (B) SIO-induced magnetic anisotropy energy ($K_{in}$ >0), which favors the ⟨100⟩ easy axis. (C) Effective magnetic anisotropy energy of SL31, in which the $K_{in}$ overcomes $K_c$. Therefore SL31 shows π/4 shift of easy axis compared to LSMO. (D) Schematic diagram of the anisotropy contributions. The SIO-induced anisotropy is determined by the interfacial exchange coupling to the emergent weak ferromagnetism in SIO (green arrows), which effectively shifts the magnetic easy axis of LSMO by π/4. (E) Thickness evolution of the MA of the superlattice and the emergent magnetism in the strong SOC SIO (FM: ferromagnetic, PM: paramagnetic).

# Figure 1

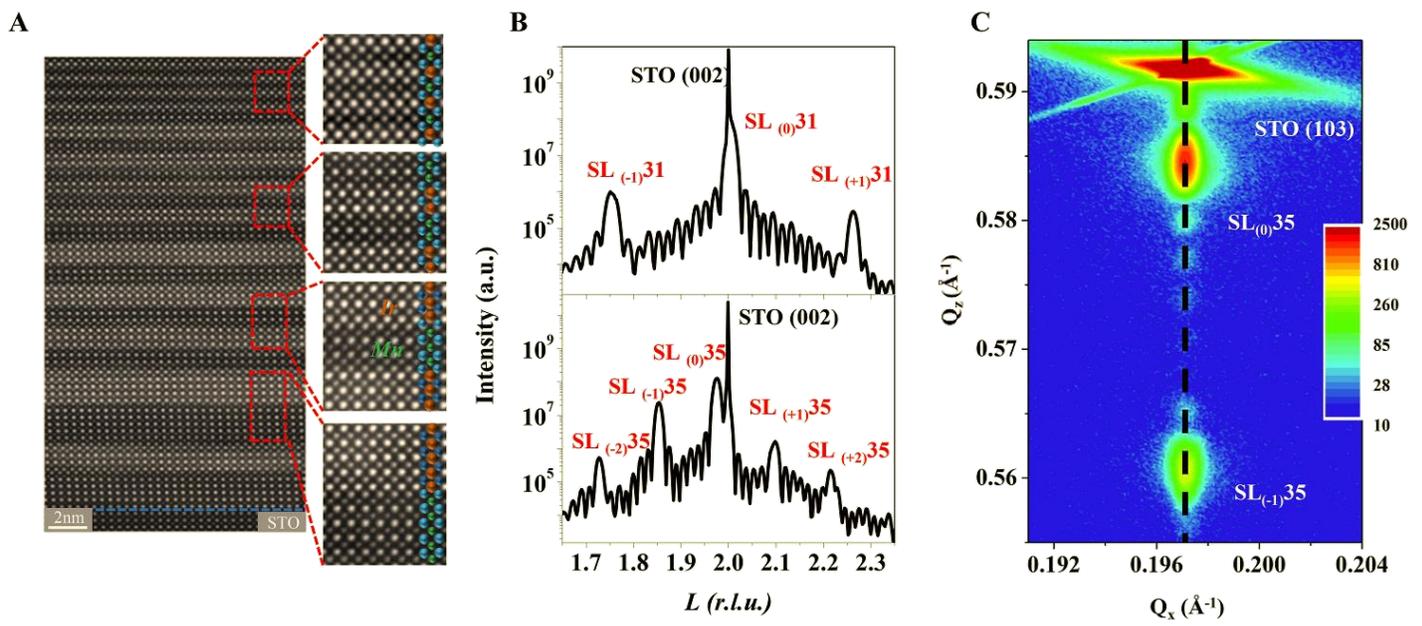

**Figure 2**

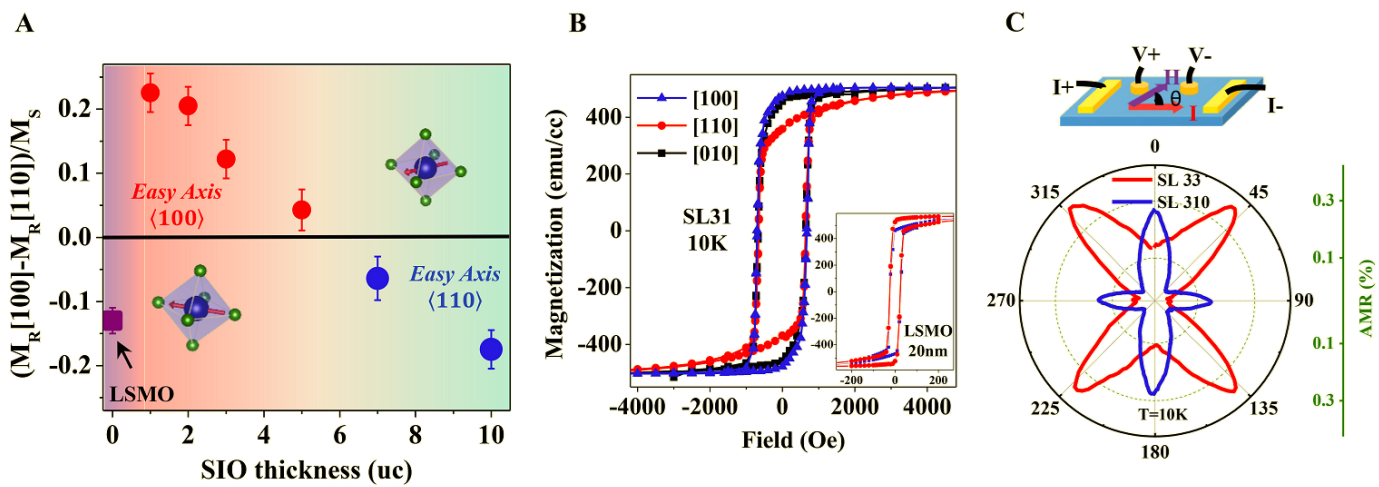

**Figure 3**

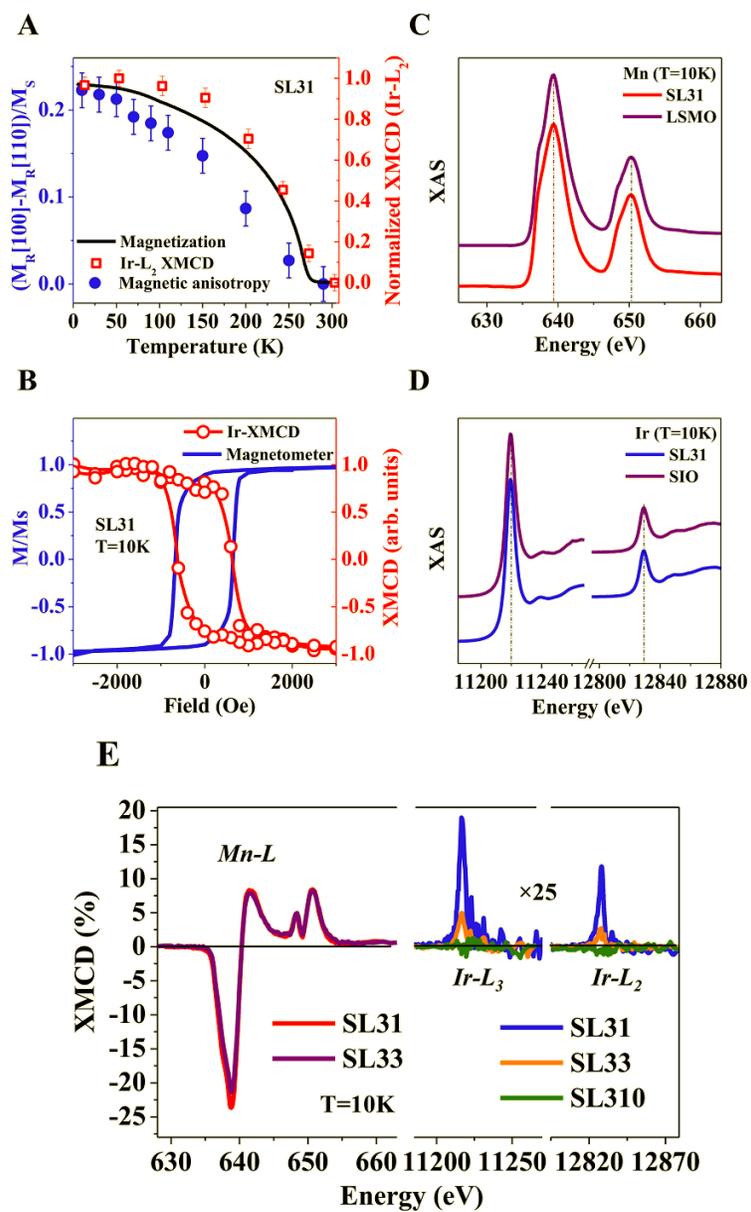

**Figure 4**

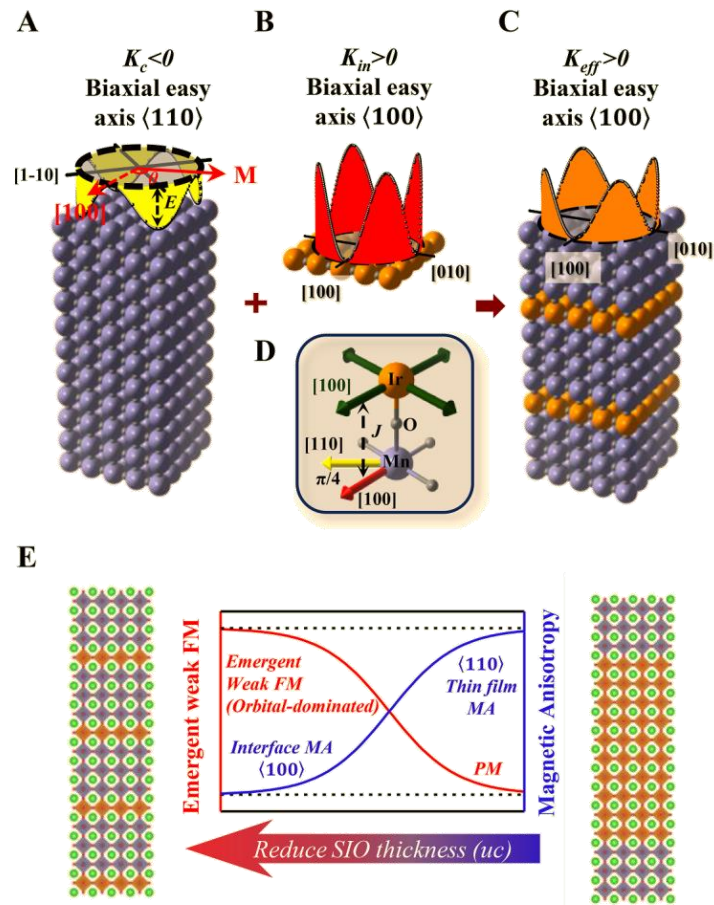

## Supplementary information:

**a. Sample preparation and structural characterization (Fig. S1 and S2)**

Fig. S1 (a) shows the RHEED patterns at different growth stages. All of the sublayers exhibit 2D RHEED patterns. Fig. S1 (b) shows the RHEED oscillations of SL33, a typical example for the series SL3$m$. The RHEED oscillations indicate the layer-by-layer growth mode of both LSMO and SIO, the key factor to fabricate high-quality superlattice. The total thickness of the superlattices is kept around 20-25nm (e.g. [SL31]$_{13}$, [SL32]$_{10}$, [SL33]$_{10}$, [SL35]$_{8}$, [SL37]$_{6}$, [SL310]$_{6}$, the number refers to the repetitions). Fig. S1 (c) is the atomic force microscopy (AFM) image of SL33, which clearly shows the unit-cell step terraces with the width of 600-800 nm, identical to that of the low-miscut TiO$_2$-terminated STO substrate. All the superlattices in the SL3$m$ series show the similar topology as Fig. S1 (c). The width of the steps and the consistent orientation of the steps of the substrates used for the SL3m also indicate that the newly observed anisotropy change described in the main text doesn't originate from step-terrace structure (8). Furthermore, the shape anisotropy induced from the step-terrace structure is uniaxial while the observed anisotropy is biaxial. Fig. S1 (d) shows the θ-2θ X-ray diffraction measurements of the superlattice series SL3$m$. The satellite peaks correspond to the periodicity and the thickness fringes reveal the high-quality of the samples.

Fig. S2 shows the STEM image and electron energy loss spectroscopy (EELS) of La M-edge across the interface in the area (LSMO)$_5$(SIO)$_5$. The STEM images shown in this study were filtered by Average Background Subtraction filter (ABSF) to reduce noise. The Z-contrast (Fig. S2 (a)) is dominated by the large difference of atomic number (Z) between B-site atoms Mn and Ir. Therefore the EELS line profile of La across the interface is used to check the A-site interfusion. Fig. S2 (b) shows the EELS line profile of the (LSMO)$_5$(SIO)$_5$ region. The blue and red background correspond to the LSMO and SIO sublayers identified by the Z-contrast, which match with the EELS line profile. The results suggest the A-site inter-diffusion is within half unit cell at the interface.

## b. Octahedral rotation pattern of superlattices (Fig. S3)

Previous study has revealed crystal symmetry change may lead to a reorientation of magnetic easy axis of LSMO from ⟨110⟩ to ⟨100⟩. As demonstrated by Boschker and coauthors (5), LSMO thin films under moderate compressive strain have the orthorhombic structure with [001]$_o$ in the film plane (larger compressive strain would change the easy axis to the out-of-plane direction). The phenomenon is suggested to arise from the asymmetry of octahedral rotation pattern along two in-plane directions of the orthorhombic distorted LSMO, which leads to a different degree of overlap of $d$ orbitals. The superlattices in this study are coherently strained, confirming that LSMO is under tensile strain, same as LSMO thin film on STO substrate. On the other hand, SIO thin film is orthorhombic with [001]$_o$ in the film plane. It has been demonstrated that octahedral rotation pattern could be altered due to interfacial octahedral mismatch (30). Therefore we examined the possibility whether the LSMO layer in the superlattice has the orthorhombic distortion with [001]$_o$ in the film plane due to the SIO adjacent layer. Here we concentrate on SL31 (the one shows the strongest MA with <100> magnetic easy axis) and SL310.

In perovskite oxides, octahedral rotations offset the oxygen atoms from the face-centered positions effectively doubling the pseudocubic unit cell, producing a distinctive set of half-order Bragg peaks depending on the octahedral rotations pattern (41). To describe the phase of the octahedral rotations along each axis, we employ the Glazer notation, in which a superscript is appended to each axis to indicate whether neighboring octahedral rotate is in-phase (+), out-of-phase (-) or no rotation (0). For example, rhombohedral unit cell shows the $a^-a^-a^-$ rotation pattern and orthorhombic unit cell shows the $a^+b^-c^-$ with [001]$_o$ along a axis. The method to determine the rotation pattern is illustrated in details in the reference (41), which is based on presence or absence of specific half-ordering peaks. The rotation patterns can generally be identified by measuring a series of [H/2, K/2, L/2] reflections. Therefore there are three types of half ordering peaks: peaks with one half-integer and two integers, peaks with two half-integers and one integer and peaks with three half-integers. Peaks with three half-integers correspond to the rhombohedral structure ($a^-a^-a^-$, out-of-phase rotation along all directions). Peaks with one half-integer and two half-integers both correspond to the orthorhombic structure ($a^+b^-c^-$, in-phase rotation along a and out-of-phase along b, c).

To examine possible orthorhombic distortion of LSMO as the origin of MA engineering, we need to check the two classes of peaks corresponding to different rotation patterns. Each half ordering reflection has a different structure factor and thus a different intensity. Zhai and coauthors have carefully measured 50 different half-ordering peaks on manganite system, which provides a solid reference to differentiate the rotation patterns (42). Here we chose the peaks with high intensity in our study. Specifically we measured the (0.5, 0.5, 1.5), (1.5, 2.5, 2.5) and (0, 0.5, 3), (0.5, 0, 3). If LSMO develops the orthorhombic structure due to the SIO, (0, 0.5, 3) and (0.5, 0, 3) peaks should be observed. Fig. S3 shows the X-ray diffraction measurements of the half-ordering peaks of SL31 and SL310 taken at beamline 6-ID-B and 33BM, APS in Argonne National. As for SL31 (blue), the three half-integer peaks are observed (Fig. S3 (a) and (b)) and the one half-integer peaks are absent (Fig. S3 (c) and (d)), which can be explained by the octahedral rotation without any in-phase rotations. Thus it should be $a^-a^-c^-$ considering the biaxial strain. This results therefore rule out the mechanism suggested by Boschker and coauthors (5) as the origin of the MA engineering observed here. For SL310 (red), the one half-integer peaks are clearly observed (Fig. S3 (c) and (d)), which is consistent with the orthorhombic structure of SIO ($a^+b^-c^-$). In conclusion, the measurements qualitatively rule out the change of structure of LSMO as the origin of MA engineering by considering the absence of peaks corresponding to orthorhombic distortion.

c. **Magnetization measurements (Fig. S4)**

Fig. S4 (a) shows the temperature dependence of magnetization of the superlattice series SL3*m*. The curve was measured during warming with 200 Oe field applied along the [100] after 1T field cooling. The superlattices show a decrease of $T_c$ as $m$ increases, similar to what has been observed in the LSMO/STO superlattices (43, 44). The effect is probably related to the interlayer coupling of LSMO (43) or the emergent ferromagnetism in the spacer layer (44), both of which depend on the spacer layer thickness.

Fig. S4 (b) shows the different crystallographic directions along which the magnetic loops were measured (defined in the pseudo-cubic notion). Fig. S4 (c) shows the magnetization loops of SL31 along in-plane [100] and out-of-plane [001] directions, revealing that the easy axis of SL31 is in the film plane. Fig. S4 (d)-(h) show the magnetic loops along two in-plane axis (black for [100] and red for [110]) for SL3$m$ ($m$=2,3,5,7 and 10), from which the thickness dependence of MA in Fig. 2A is extracted. The $\pi/4$ shift of magnetic easy axis is clearly demonstrated. All the hysteresis loops were taken at 10K and the magnetization were averaged by the LSMO thickness.

### d. Transport measurements (Fig. S5)

Fig. S5 (a) shows the temperature dependence of resistance (normalized by the resistance at 300K) for the superlattice series SL3$m$. As $m$ increases from 1 to 10, the temperature dependence behavior changes from LSMO-like to SIO-like. Fig. S5 (b) shows the AMR of 4uc SIO thin films as a reference sample. The black curve is the AMR measured with 1T magnetic field, which shows the magnitude of 0.01% (close to the resolution of the set-up). The amplitude of AMR of the 4uc SIO increases with the magnitude of magnetic field. AMR at 14T shows 4-fold rotation dependence with the amplitude of 0.1% and minimum at 45 degree ([110]) instead of 90 degree [100] (the definition of angle is the same as main text). The amplitude of AMR for the superlattices measured with 1T is in the order of 1%, which is one order higher than that of SIO shown in Fig. S5 (b). Therefore the results suggest that the AMR signal is mainly originated from the colossal magnetoresistive LSMO, which is strongly coupled with SIO.

Fig. S5 (c) shows the AMR measured at 10K with 1T field rotated in-plane for SL3$m$. The measurement geometry is the same as main text (Fig. 2C). For SL31, the AMR is dominated by the uniaxial component

(2-fold rotation dependence), which is commonly observed in ferromagnetic metals and independent on the lattice symmetry. As the thickness *m* increases, the biaxial anisotropy (four-fold) component gradually dominates the AMR. Despite ongoing debate on the physical origin, the 2-fold and 4-fold AMR has been observed and discussed in manganite systems before (45), establishing several experimental facts. First, the four-fold AMR reflects the magnetocrystalline anisotropy of the manganite system (45). Second, there is a four-fold to two-fold AMR transition as the applied magnetic field increases. The transition due to the magnetic field was also observed in the superlattice SL33 as shown in Fig. S5 (d), which again confirms the dominant contribution from LSMO. The phase of the AMR (the angle of the maximum/minimum) has an interesting thickness dependence, which shows the $\pi/4$ shift by comparing *m*=2,3,5 with *m*=10. The results are consistent with the magnetic easy axis evolution shown in Fig. 2A and therefore confirm the MA evolution. Fig. S5 (e) shows the temperature dependence of AMR for SL33. The $\pi/4$ phase shift of AMR (maximum at 45° instead of 90°) persists to high temperature around Curie temperature, which is consistent with the magnetization measurement in Fig. 3A.

e. **XMCD characterization and analysis (Fig S6, S7)**

The sum-rules is applied as following (46) to both the Mn and Ir *L* edge to analyze the XMCD spectra:

$$m_{orb} = -\frac{4\int_{(L_2+L_3)}(\mu_+ - \mu_-)\,dE}{3\int_{(L_2+L_3)}(\mu_+ + \mu_-)\,dE}(10 - n_{3d})$$

$$m_{se} = -\frac{6\int_{(L_3)}(\mu_+ - \mu_-)\,dE - 4\int_{(L_3+L_2)}(\mu_+ - \mu_-)\,dE}{\int_{(L_2+L_3)}(\mu_+ + \mu_-)\,dE}(10 - n_{3d})$$

Fig. S6(a) shows the schematic on the application of the sum-rules for XMCD at Ir-edge. For the hard X-ray XMCD, the normalization is conventionally performed by setting the edge jump of $L_3$ as 1 (38). For the soft X-ray XMCD, the normalization is conventionally performed by setting the maximum of XAS as 1. However no matter which normalization method is applied, the results yielded by sum-rules are normalization-independent because the XAS and XMCD are multiplied by the same factor during the normalization process. In this study, the Ir-XMCD here is normalized by the maximum of $L_3$ XAS as shown in Fig. S6(a). In the normalization process, the ratio of $L_3$ edge jump to $L_2$ edge jump is kept as 2. In Fig. S6(a), the black dash line is the background absorption spectra expressed by the two step functions. The blue shadow areas correspond to the integration of $L_2$ and $L_3$ XAS: ($\int_{(L_2)} (\mu_+ + \mu_-)/2 \, dE$ and $\int_{(L_3)} (\mu_+ + \mu_-)/2 \, dE$). The red shadow areas correspond to the integration of $L_2$ and $L_3$ XMCD ($\int_{(L_2)} (\mu_+ - \mu_-) dE$ and $\int_{(L_3)} (\mu_+ - \mu_-) \, dE$).

Fig. S6 (b)-(g) shows the integration ("Sum" in each figure) of XAS and XMCD for Ir $L_3$ and $L_2$ edges of both the SL31 and SL33. By using the values of integration (Sum) in the formula of sum-rules, the orbital angular momentum and the effective spin momentum can be obtained. For SL31, the analysis yields $m_l = (0.036 \pm 0.003) \, \mu_b/Ir$ and $m_{se} = (0.002 \pm 0.003) \, \mu_b/Ir$. (Error bar is estimated by performing the sum-rule analysis on multiple measurements) The isotropic spin component cannot be extracted from the effective spin momentum without a priori knowledge of the value of the $\langle T_z \rangle$ ($m_{se} = m_s + 7\langle T_z \rangle$, $\langle T_z \rangle$ is the magnetic dipole operator), which requires further investigations. Besides, it is not always accurate to correlate the value of $m_l$ and $m_s$ from the sum-rule analysis to that of the real system, which is primarily because the absolute number of holes within the valence shell is difficult to extract from x-ray absorption. Nevertheless, one can still compare reliably the obtained ratio between the orbital and effective spin momenta with that of an ideal $J_{eff}=1/2$ state (38).

Moreover, the Ir edge XMCD spectra show an interesting thickness dependence. As shown in Fig. 3E, the XMCD signal reduces as *m* increases. A careful analysis of the results suggests that the reduction cannot be simply attributed to the decreased interface density within the superlattices. First of all, no XMCD signal was observed on SL310 as shown in Fig. 3E, which would otherwise show a small but observable signal. Second, the comparison between SL31 and SL33 XMCD further supports the conclusion. As shown in Fig. S6 (d) and (g), the XMCD signal of SL33 is roughly 1/4 of that of SL31 on $L_3$ edge and 1/5 on $L_2$ edge, neither one of which scales with the total thickness. More interestingly, $L_2/L_3$ ratio reduces as m increases, which would lead to a change of $m_l/m_{se}$ by applying the sum-rules. All these results show the SIO thickness dependence of the emergent magnetism and new spin-orbit state. The effect of thickness confinement on SIO has also been shown in other systems (22, 25), albeit without the novel spin-orbit state observed here. The possible origin of the effect could be the change of nearest-neighboring coupling or the change of octahedral rotation of SIO (27) as the *m* increases.

The sum-rule analysis also suggests that the novel spin-orbit state is likely to be a mixture of $J_{eff}=1/2$ and $J_{eff}=3/2$. $J_{eff}=1/2$ originates from the atomic $J_{eff}=5/2$ with same sign of orbital and spin moment. The increase of $m_l/m_s$, observed by XMCD, suggests that $J_{eff}=1/2$ is likely to mix with another state that has the opposite sign of orbital and spin moment, i.e. the $J_{eff}=3/2$ ($J_{eff}=3/2$ originates from the atomic $J_{eff}=3/2$ with opposite sign of orbital and spin moment). The mixture is possible if the degeneracy of $t_{2g}$ orbitals is lifted, for example, by the interfacial orbital reconstruction (in the ideal $J_{eff}=1/2$, the ratio of xy:yz:xz is 1:1:1). Alternately, the mixture could also arise from the interface magnetic coupling, which will be discussed in SI Appendix section g.

Fig. S7 show the integration ("Sum" in each figure) of XAS and XMCD of Mn edge for the SL31. The sum-rules tends to underestimate the magnetic moment when spin-orbit splitting of the transition-metal 2*p* core levels is not large enough compared to the 2*p*-3*d* exchange splitting interaction, leading to uncertainties in the estimation process for the low 3*d* TMOs (47). However by considering the geometry factor of the measurement (incident angle) and the 'correction factor' given in the reference (42), the effective spin component of the Mn moment here is estimated to be around 2.7 $\mu_b/Ir$ and the orbital component is estimated to be less than 0.02 $\mu_b/Ir$. Unlike SIO, the $m_l/m_{se}$ is drastically smaller, consistent with the dominant role of the spin moment for *3d* TMOs. The calculated value is consistent with our macroscopic SQUID measurement.

**f. First principle calculations (Fig. S8 and Table S1)**

Density functional theory calculations were carried out using the Vienna Ab-Initio Simulation Package (48) (VASP) with the projector augmented wave (PAW) method. The corresponding electronic configurations for each element are Sr: 4s4p5s; Ir: 5d6s; O: 2s2p; La: 5s5p6s5d; Mn: 3d4s. The cutoff energy is chosen to be 550 eV based on the convergence tests. The revised version Perdew-Burke-Ernzerhof functional for solids (PBEsol) under the generalized gradient approximation (GGA) was chosen as the exchange-correlation function. Spin-orbital coupling is included in the simulations. For the spin-orbit coupling (SOC) calculation, unconstrained noncollinear magnetism (49) setting are employed. The quantization axis for spin was set as [0 0 1]. The local magnetic moment of Ir and Mn atoms in x, y, and z directions was initialized and then subsequently relaxed. In addition, the rotationally invariant+$U$ method (50) introduced by Liechtensitein et al. were applied. The chosen effective coulomb-$U$ parameters for each element are discussed in the following part. We used 4×4×2 K-points following the Monkhorst-Pack scheme in our systems. The convergence criterion for the electronic relaxation is $10^{-6}$ eV.

To simulate the SL31 superlattice, we worked with the basic cell depicted in Fig. S8 (a), which comprises 4 perovskite layers in the (001) orientation with totally 20 atoms. In periodic calculations it is not possible to properly mimic the ionic disorder La/Sr. In this calculation, we treated it in an ordered way by keeping the ratio as 2/1. When relaxing the cell parameters and atomic positions, the in-plane lattice constant (a) is constrained to that of STO and we make sure that c/a fits the experiments.

We used different combinations of U values for Mn, Ir and La in our U-corrected DFT approach to the LSMO/SIO superlattice as shown in Table S1. The Coulomb interaction parameter U for Ir $5d$ orbital was set to be 2 eV, which was successfully used in calculation of perovskite Iridates (51). The U of Mn $3d$ orbital was chosen to be 3 ev or 4eV, commonly used for the manganites (18). For La, we considered La without U and with U=3 eV on 5d orbital. All of the parameters yield similar structure as shown in Fig. S8(a). The rotation pattern is out-of-phase along the c direction (c⁻), consistent with the experimental results shown in Fig. S3. In order to determine the magnetic easy axis, the magnetic moment of Mn was set to be along the pseudo-cubic [100] or [110] and the magnetic moment of Ir was free to relax in each case to minimize the energy. Then the obtained energy minimum of the two configurations were compared to determine the magnetic easy axis (lower energy). All of the parameters in Table S1 yield equivalent result with ⟨100⟩ magnetic easy axis despite of slight numerical differences in the energy. Therefore the calculations indicate that the shift of magnetic easy axis to ⟨100⟩ in SL31 is a robust feature.

Fig. S8 (b) shows the relaxed magnetic structure of the low-energy state (⟨100⟩ easy axis) obtained from the calculation. The moment of each Mn cation aligns along [$\bar{1}$00] direction (~3.4 $u_b$/Mn with dominant spin component). In the SIO monolayer, an in-pane canted antiferromagnetic ordering (weak ferromagnetism) emerges with the moment of each Ir cation along [010] direction (perpendicular to the moment of Mn cation). The canting of Ir moments yields a net moment (~0.03 $u_b$/Ir) that is antiparallel to moment of Mn cation, consistent with the experiment result. This magnetic structure is consistent with the "spin-flop" mechanism observed at many ferromagnet/antiferromagnet interfaces.

To further validate the calculation, we also carried out the same calculation on LSMO without the SIO monolayer by using the same parameters. The spin direction of Mn is set to be along [100] and [110] initially. After the spin-orbit calculation, the energy of each relaxed magnetic state is obtained as shown in table S1. The result suggests the easy axis along [110], consistent with experiment, although the energy difference is much smaller compared with SL31. The significant reduction of energy difference is due to the weak SOC of Mn (SOC is proportional to $Z^4$, Z is the atomic number), which is the origin of the magnetocrystalline anisotropy. In general, it is a difficult task to calculate the magnetic anisotropy energy of 3d TMOs by first-principle since the energy difference in many cases are extremely small (52). The references on this topic are very limited. To our best knowledge, we cannot find other reference that calculates the magnetic anisotropy energy of manganite. However calculation results on other 3d TMOs yield the energy difference in the similar order of magnitude (53). Therefore our first-principle calculation results qualitatively support the trend we observed experimentally.

### g. Further discussions of the magnetic anisotropy (Fig. S9)

Insights into the interface-controlled magnetic anisotropy observed in the superlattices can be gained through comparison with the long-range magnetic orderings in $Sr_2IrO_4$ and $Sr_3Ir_2O_7$. $Sr_2IrO_4$ shows the in-plane canted antiferromagnetic ordering with moment axis ⟨110⟩ (35, 36). $Sr_3Ir_2O_7$ shows the c-axis collinear antiferromagnetic ordering where moment axis [001] without canting (40). Our experimental results suggest that, while the dimensional magnetic evolution is different from that of the RP series, the emergent magnetic ordering is likely to be the canted in-plane antiferromagnetism (like $Sr_2IrO_4$). However the magnetic anisotropy of the superlattice, combined with ab-initio calculation, suggests a ⟨100⟩ easy axis for Ir moments, different from ⟨110⟩ in $Sr_2IrO_4$. The mechanism of the ⟨110⟩ easy axis of $Sr_2IrO_4$ has been discussed by considering the competition between quantum fluctuations and interlayer coupling (35, 36). This model suggests that, although moments of a 2D $IrO_2$ plane prefers to align along ⟨100⟩ directions, the interlayer coupling lowers the energy when moments align along ⟨110⟩ due to the layer-stacking sequence of $Sr_2IrO_4$(214) phase (Ir cations shift by [0.5, 0.5] between neighboring $IrO_2$ planes). In contrast, in the superlattice, the interlayer coupling between $IrO_2$ planes is weak since the two adjacent layers are separated by 3uc LSMO. More importantly, the stacking with a perovskite-type superlattice renders a crystal symmetry completely different from 214 phase. Therefore, applying the same theoretical model would conclude that the moments of Ir in the superlattice geometry should align along ⟨100⟩ instead of ⟨110⟩. However, a shortcoming of this model is that it is only applicable to the single $IrO_2$ plane limit and the energy scale of this anisotropy would be rather weak due to limitation of the quantum zero-point energy.

Notice that all the models applied to the RP phases are based on the validity of $J_{eff}=1/2$ state as a good approximation. As we discussed above, the new spin-orbit state in the superlattice is significantly deviated and likely to be a mixture of $J_{eff}=1/2$ and $J_{eff}=3/2$. Therefore one must consider the magnetic anisotropy contributed by the mixed $J_{eff}=3/2$ component, which, unlike $J_{eff}=1/2$, has a relatively large single-ion anisotropy as discussed in the following.

To understand the origin of the anisotropy, one can write down the Hamiltonian matrix that includes the spin-orbit coupling energy ($\lambda$), cubic crystal field energy (10Dq) and magnetic energy (h, an effective magnetic field coupled to the onsite moment along the quantization axis which is set to be along the Z axis) in the cubic limit. The stability of a certain magnetic moment direction, i.e. single-ion anisotropy, is ultimately determined by the size of the Zeeman splitting. The matrix can be reorganized into four independent blocks with the corresponding ten states in d-shell as shown in Fig. S9(a). There are clearly off-diagonal matrix elements due to SOC and Zeeman splitting that mix different states. The presence of these matrix elements is due to the fact that these "ideal" states are *not* eigenstates of the SOC operator or the angular momentum operators.

The first block reveals a hybridization between $J_{eff}=3/2(-3/2)$, $J_{eff}=1/2(-1/2)$ and $e_g(x^2-y^2$, up) orbitals by the off-diagonal matrix elements. Based on the perturbation theory, the second order correction to eigenenergy of $J_{eff}=1/2$ is proportional to $h^2/\lambda$ (hybridization with $J_{eff}=3/2$) and $h^2/10Dq$ (hybridization with $x^2-y^2$). Both are purely due to Zeeman splitting. On the other hand, the change of eigenenergy of

$J_{eff}=3/2$ is proportional to $(h+\lambda)^2/10Dq$ (with $x^2-y^2$). For iridates, the crystal field 10Dq is round 3 eV (54) and SOC energy $\lambda$ is about 0.4 eV (37). It is not easy to evaluate h. However, even if we treat it as a large effective exchange field (e.g. because of the magnetic coupling of Ir-Mn), a reasonable estimation could only be as large as the order of 10 meV, which leads to 10Dq, $\lambda \gg h$. Thus the correction is small for $J_{eff}=1/2$. In other words, the $J_{eff}=1/2$ state has no significant single-ion magnetic anisotropy since the energy is not sensitive to the orientation of moment with respect to the crystal field. On the other hand for $J_{eff}=3/2$ state, because $h*\lambda/10Dq \gg h^2/10Dq$, the contribution from any effective exchange field would be effectively amplified by a factor of $\sim \lambda/h$ through the hybridization with the $e_g$ ($x^2-y^2$) state. Meanwhile, since this effect relies on the mixing with the $e_g$ ($x^2-y^2$) state on the order of $\lambda^2/10Dq$, it is maximized when the Zeeman quantization axis is aligned with a crystal field principle axis. In other words, the $J_{eff}=3/2$ should have a large single-ion anisotropy since the energy is sensitive to the orientation of moment with respect to the crystal field.

To demonstrate this effect, we also simulated the angle dependence (in the XZ plane) of Zeeman splitting energy ("up"-state minus "down"-state) by replacing Zeeman term $(L_z+2S_z)h$ with $[\cos\theta(L_z+2S_z)+\sin\theta(L_x+2S_x)]h$ in the Hamiltonian. In the simulation, we set 10Dq=3, $\lambda$=0.4 and h=0.02 in the unit of eV. The simulation results are shown in Fig. S9(b). The Zeeman splitting energy of $J_{eff}=3/2$ state shows a significant angle dependence, reflecting a large single-ion anisotropy with easy axis along the crystal bond directions ($\theta=n\pi/2$, $\langle 100 \rangle$). On the other hand, Zeeman splitting energy of $J_{eff}=1/2$ shows a very weak dependence, consistent with the small single-ion anisotropy. To further demonstrate the origin of the anisotropy of the $J_{eff}=3/2$ state, we also examined situations where the SOC energy ($\lambda$) and crystal field (10Dq) are zero respectively, which reveal their essential role in determining the anisotropy. In total, the simulations suggest that the $J_{eff}=3/2$ state has a large single-ion anisotropy with easy axis $\langle 100 \rangle$, which contributes the anisotropy of the superlattices.

In sum, the emergent magnetic state of SIO in the superlattice shows both similarities and distinctions compared to the RP-phase iridates. The difference in magnetic anisotropy is likely to arise from the novel spin-orbit state of SIO in the superlattice, which is a mixture of $J_{eff}=1/2$ and $J_{eff}=3/2$. Moreover, the discussions above also reveal that magnetic coupling between LSMO and SIO at the interface could actually contribute to the mixing of $J_{eff}=1/2$ and $J_{eff}=3/2$ (see SI Appendix section e).

## SI Figure Legends

**Fig. S1** (a) RHEED patterns of STO substrate, LSMO sublayer and SIO sublayer. All of the sublayers show the 2D character. (b) The typical RHEED intensity oscillations of the LSMO (blue) and SIO (red) sublayers for SL33. (c) AFM image of SL33. The topography shows that the surface preserves the step-terrace structure of the substrate. (d) θ-2θ X-ray diffractograms of the series SL3$m$.

**Fig. S2** (a) High-angle annular dark-field (HAADF) STEM image of LSMO/SIO superlattice in (LSMO)$_5$(SIO)$_5$ periodic region. (b) EELS line profile of the La $M$-edge of the same region. The scan line is shown by the red arrow.

**Fig. S3** $L$-scan of several half-ordering peaks of the superlattice SL31 (Blue) and SL310 (red): (a) (0.5, 0.5, 1.5), (b) (1.5, 2.5, 2.5), (c) (0, 0.5, 3), (d) (0.5, 0, 3). Inset in (c) and (d) show the results of SL31.

**Fig. S4** (a) Temperature dependence of magnetization of SL3$m$. The dependence was measured during warming with 200 Oe applied in [100] after 1T field cooling. (b) Schematic of the different crystallographic directions along which the magnetic loops were measured. (c) Magnetic hysteresis loops of SL31 with field applied in-plane ([100], black) and out-of-plane ([001], green). (d)-(h) Magnetic hysteresis loops of SL3$m$ ($m$=2, 3, 5, 7, 10) along two in-plane directions: [100] (black) and [110] (red).

**Fig. S5** (a) Temperature dependence of normalized resistivity of SL3*m*. (b) AMR of 4uc SIO thin film on STO at 10K (measurement geometry is the same as Fig. 2C)). (c) Thickness dependence of the AMR for SL3*m* at 10K with 1T field applied in-plane. (d) Field dependence of AMR for SL33 at 10K. (e) Temperature dependence of AMR for SL33 with 1T field.

**Fig. S6** (a) Schematic diagram of the application of sum-rules to the Ir-edge XAS and XMCD. (b), (c), (d) show the results on *Ir-L$_3$* edge: integration ("Sum" in each figure) of (b) XAS, (c) XMCD for SL31 and (d) XMCD for SL33. (e), (f), (g) show the results on *Ir-L$_2$* edge: integration of (e) XAS, (f) XMCD for SL31 and (g) XMCD for SL33.

**Fig. S7** Application of sum-rules to the Mn L-edge of SL31.

**Fig. S8** (a) The basic cell used for calculation (black line) and the relaxed crystal structure of the SL31. (b) Relaxed magnetic structure of the SL31.

**Fig. S9** (a) The Hamiltonian Matrix that includes the spin-orbit coupling ($\lambda$), crystal field (10Dq) and magnetic energy (h). (b) Numeric simulations of angle dependence of Zeeman splitting energy (ZS, "up"-state minus "down"-state) for $J_{eff}=3/2$ and $J_{eff}=1/2$ ($\lambda=0.4$, 10Dq=3 and h=0.02 or otherwise stated). $\theta$ is the angle of moment with respect to the quantization axis Z in XZ plane ($\theta=n\pi/2$ represents the moment along the bond direction $\langle 100 \rangle$). To compare different plots, Zeeman splitting energy (ZS) is presented as ZS($\theta$)-(ZS(max)+ZS(min))/2. The maximum of ZS corresponds to the direction of easy axis. (Zeeman splitting energy of $J_{eff}=3/2$ represents the difference of eigenenergy between $J_{eff}=3/2(3/2)$ and $J_{eff}=3/2(-3/2)$).

**Table S1** Energy minimum calculated for different combinations of U parameters with magnetic moments of Mn along two crystallographic directions for SL31 and LSMO single layer

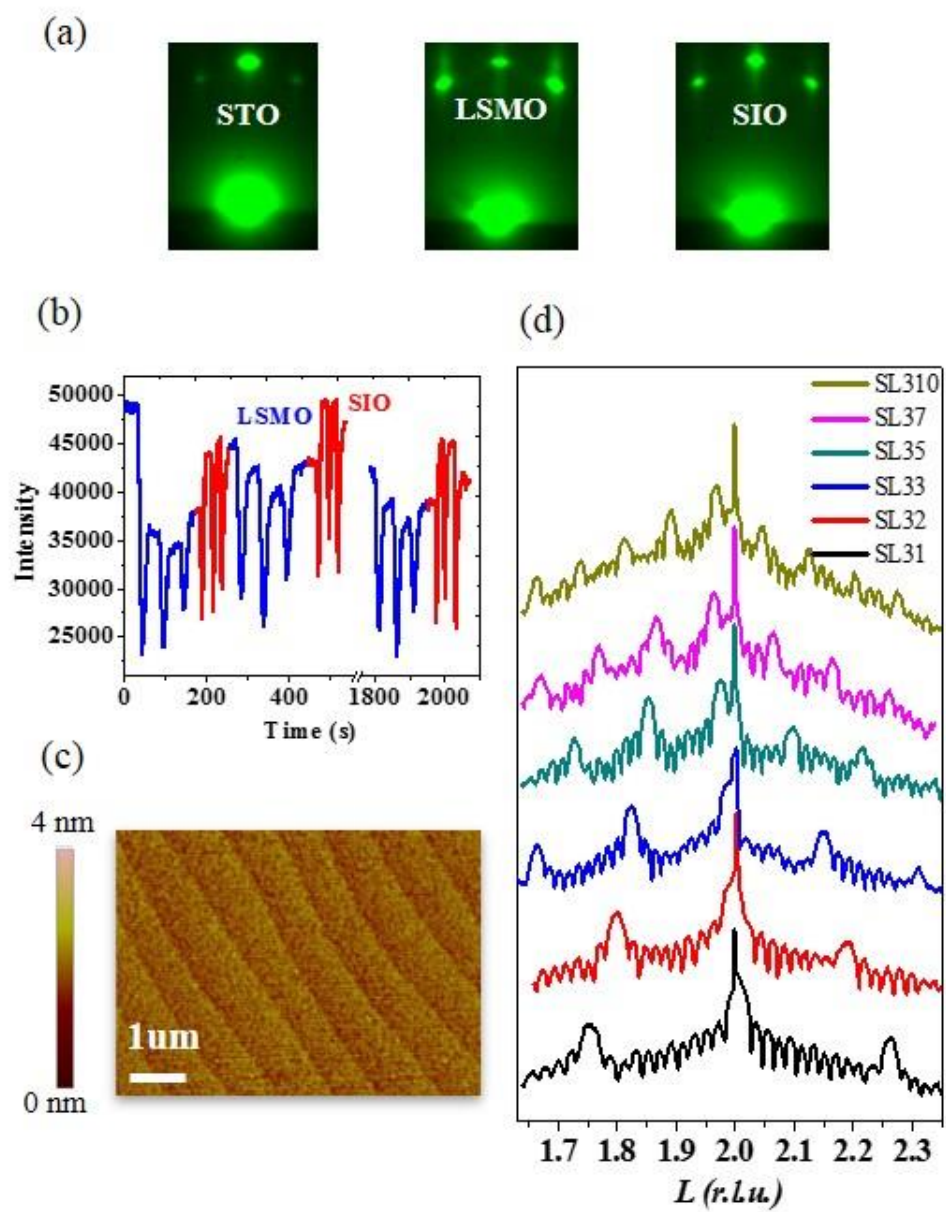

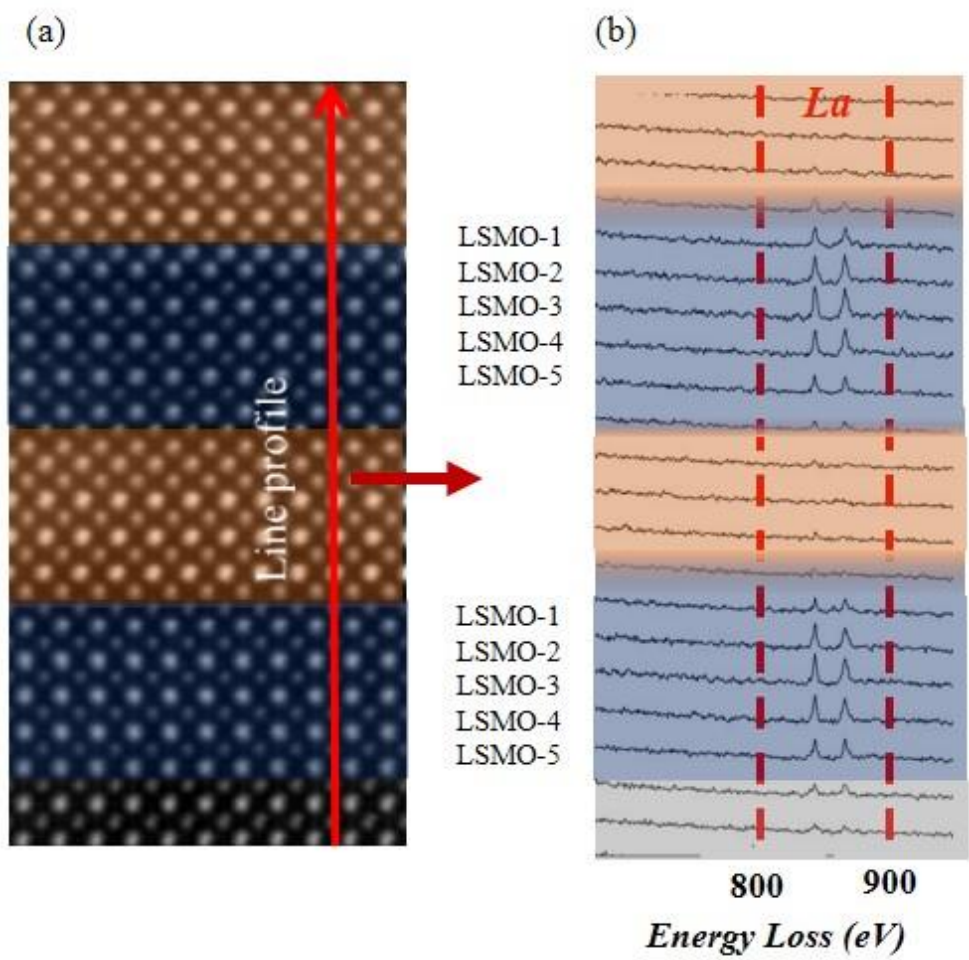

S3

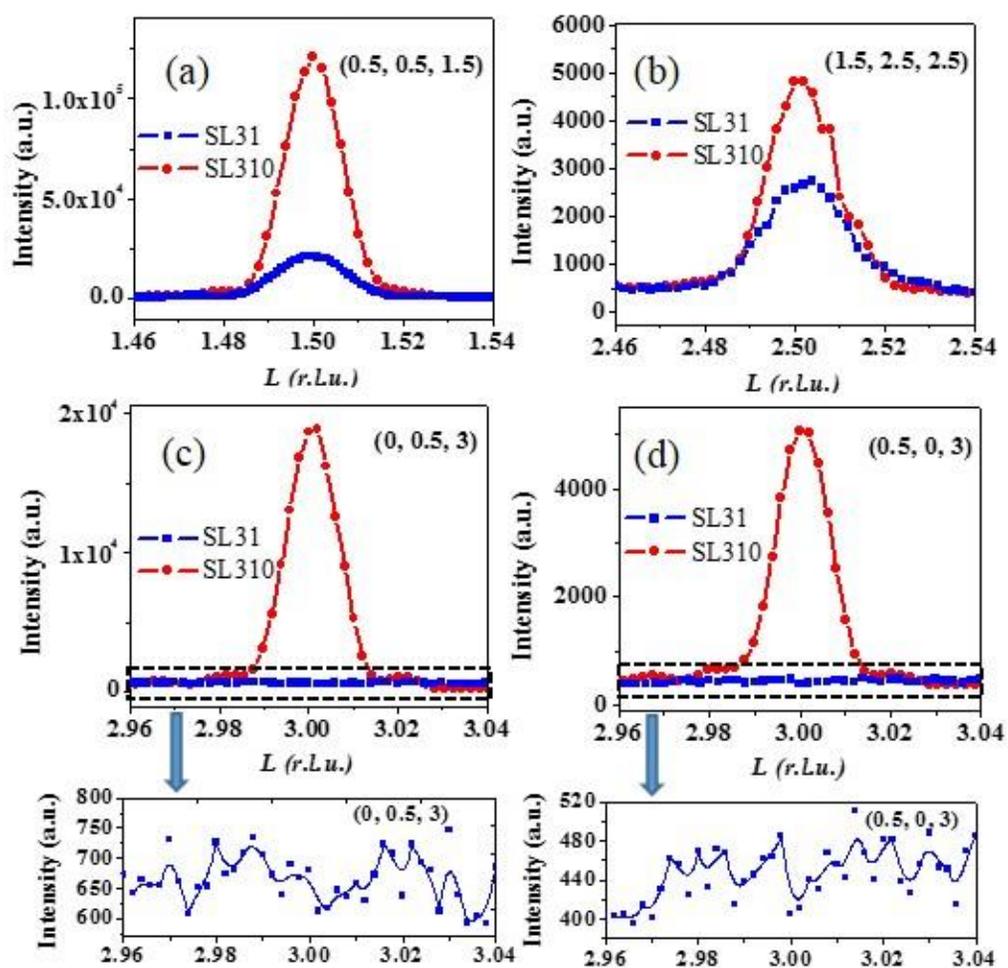

S4

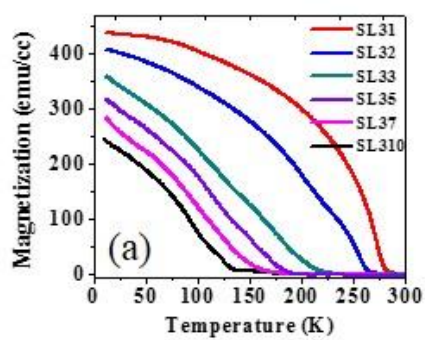
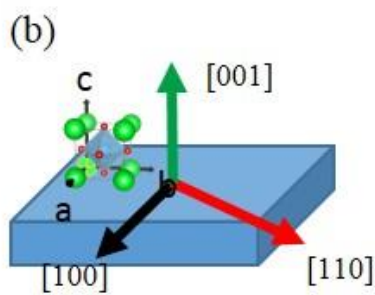
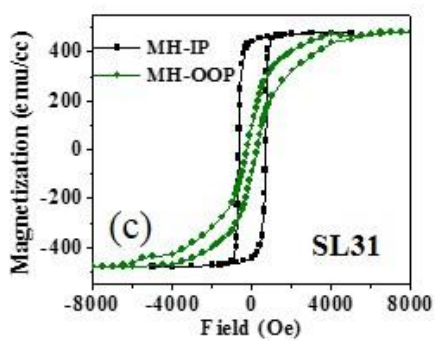
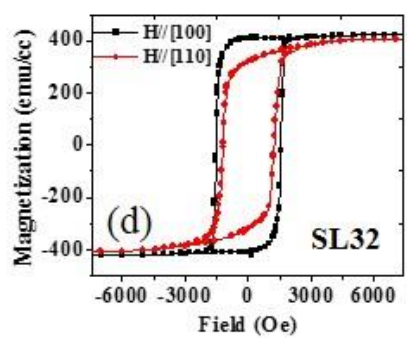
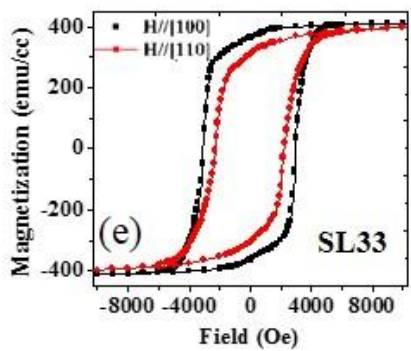
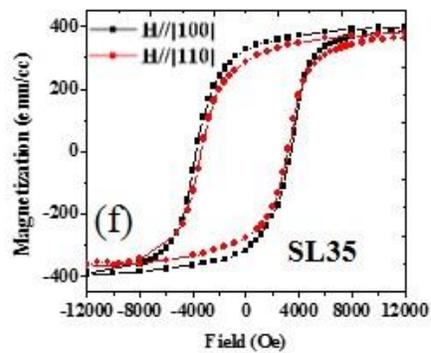
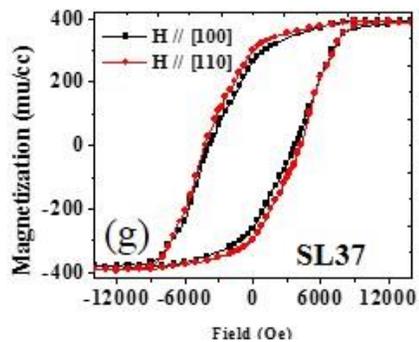
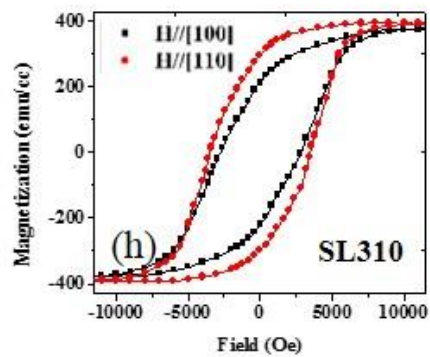

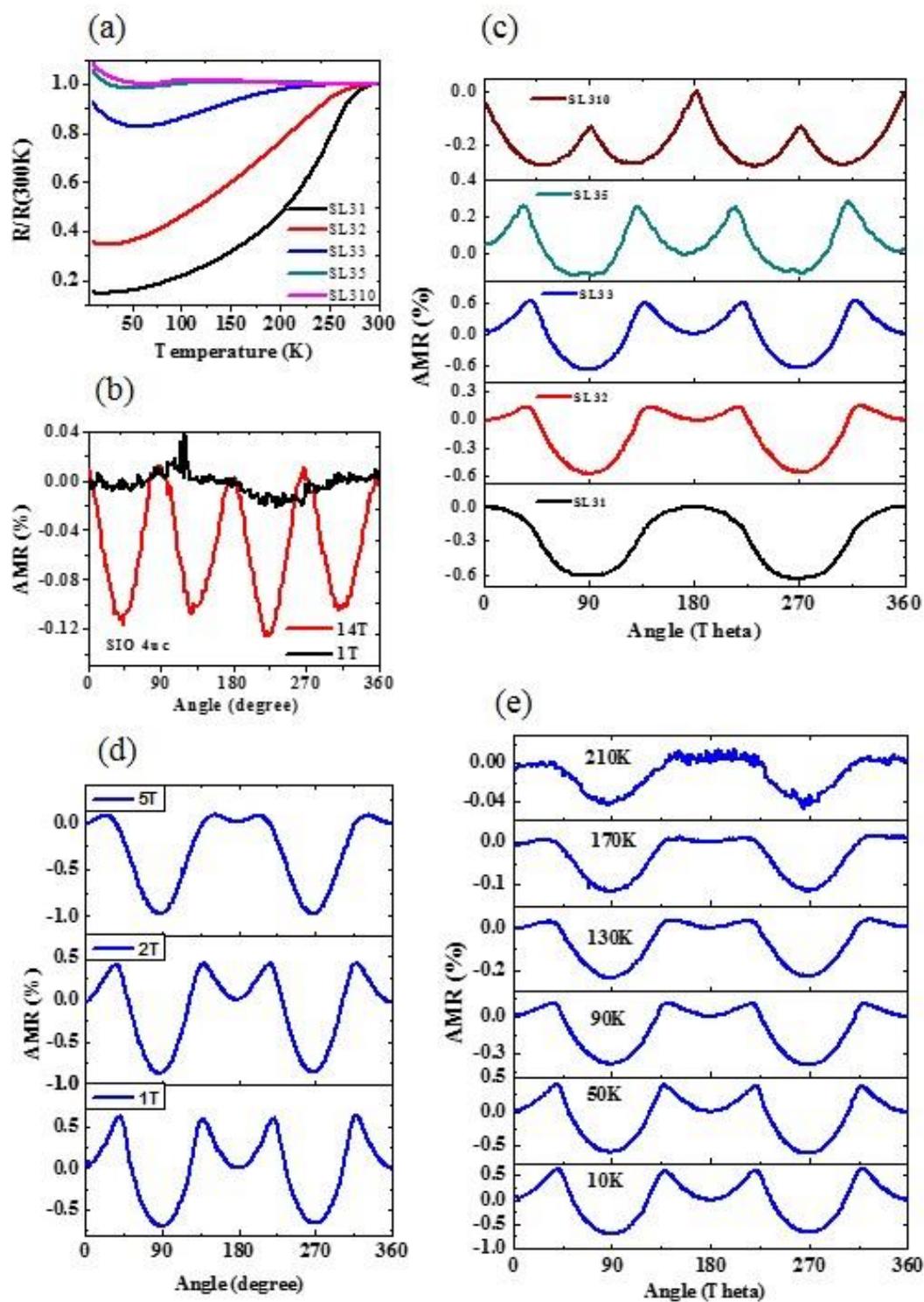

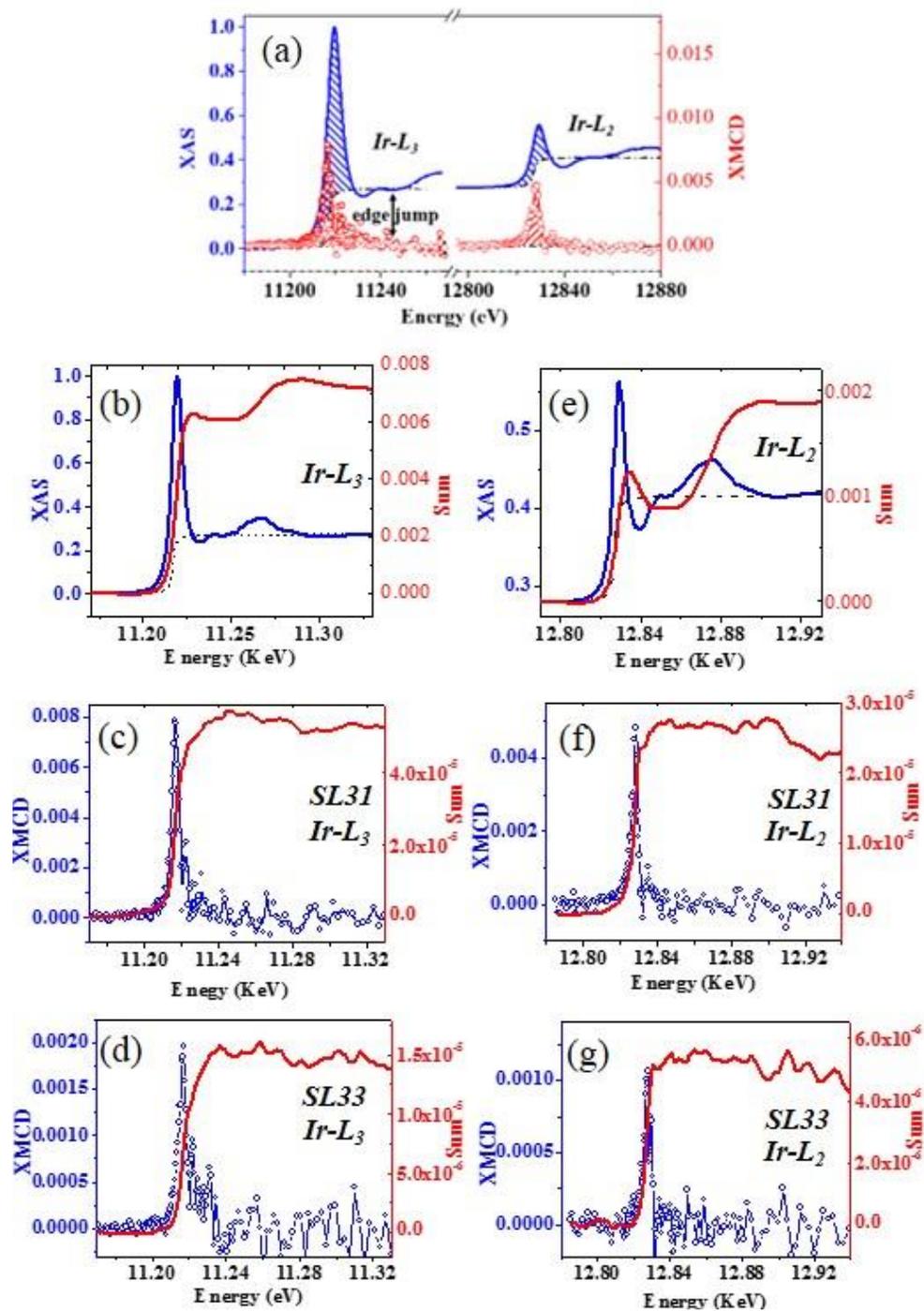

S7

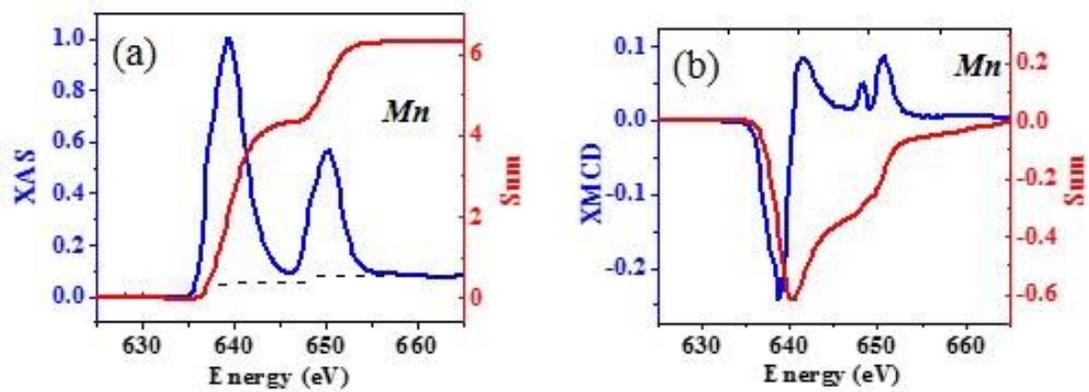

S8

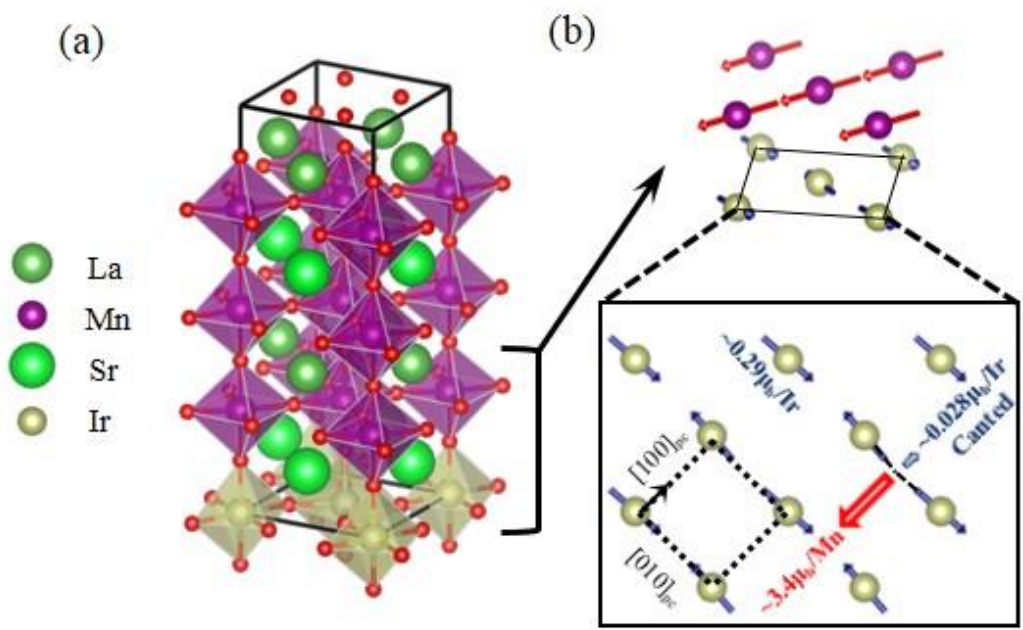

S9

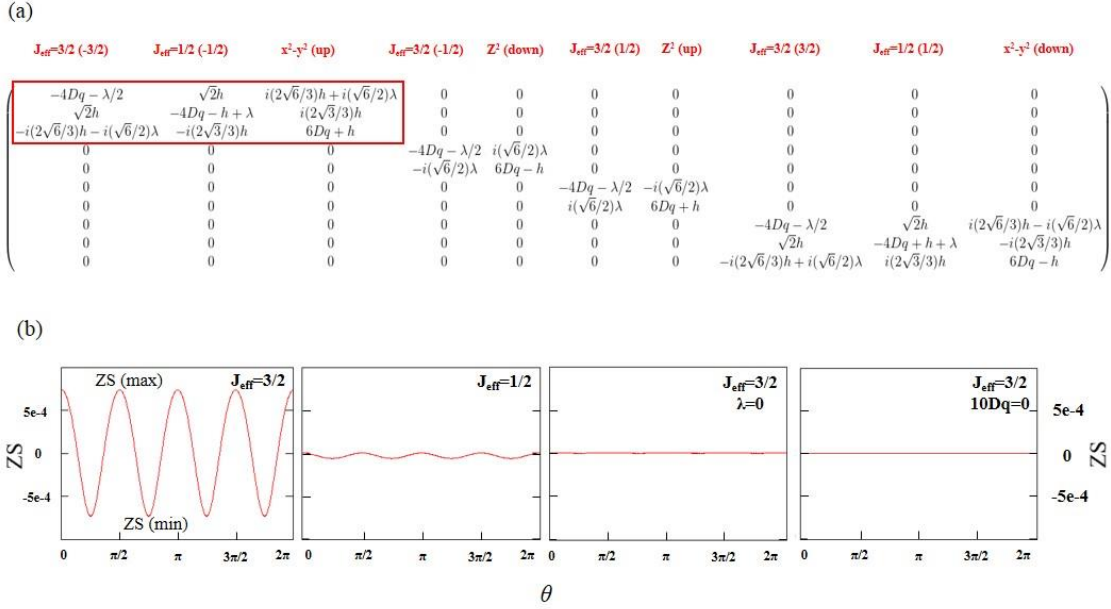

Table S1

|  | Parameters (eV) | Energy Minimum (eV) | |
|---|---|---|---|
|  |  | M // [100] | M // [110] |
| SL31 | U(Mn)=3, U(Ir)=2, U(La)=0 | -314.9843 | -314.9244 |
|  | U(Mn)=3, U(Ir)=2, U(La)=3 | -311.9687 | -311.9137 |
|  | U(Mn)=4, U(Ir)=2, U(La)=3 | -308.6973 | -308.5817 |
| LSMO | U(Mn)=4, U(La)=0 | -246.041255 | -246.041262 |
|  | U(Mn)=4, U(La)=3 | -242.999605 | -242.999608 |